\documentclass[dvips]{article}
\usepackage{icrctc07}

\begin{document}
\title{The Maximum Detectable Momentum for cosmic ray
muons in the MINOS far detector}
\shorttitle{MINOS MDM}
\authors{Maury Goodman for the MINOS collaboration}
\shortauthors{M. Goodman et al.}
\afiliations{$^1$ Argonne National Lab}
\email{maury.goodman@anl.gov}
\abstract{A magnetic detector such as MINOS which is 
measuring the sign of muons has to deal with 
issues of bending, which depend on the magnetic field configuration, 
and multiple scattering, which depends on the amount of material 
which is traversed.  Above some momentum which depends on 
these factors, the momentum cannot be resolved.  Issues related 
to measurement of the muon charge ratio in MINOS 
are discussed.
}
\maketitle

\section{Introduction}
The MINOS detector was designed to measure neutrinos produced
in a beamline at Fermilab starting with protons at the 120 GeV
Main Injector (NuMI).  The far detector is located 735 km from the
NuMI target and in the Soudan Underground Laboratory 710 m underground.
The detector consists of 486 2.5 cm octagonal
iron plates perpendicular to the NuMI beam, arranged in two 
"supermodules".  There are
484 1 cm thick scintillator planes with corresponding 2.5 cm wide air gaps.
The scintillator strips are 4.1cm wide and 3.4 - 8.0 m long.
The strips are oriented at $\pm$ 45$^\circ$ to the (vertical)
y axis, in 
orthogonal ``u"
and ``v" planes.
The steel has been magnetized in a toroidal configuration
to an average value of 1.3 T.  The field is saturated at 1.8 T near
the central coil hole and falls to 1 T near the octagon edges.  
\par
\begin{figure}[htpb]
\begin{center}
\includegraphics*[width=3cm,angle=0,clip]{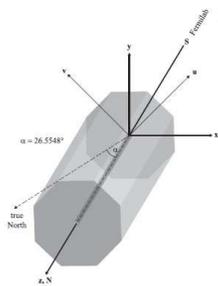}
\caption{\label {fig:det} 
The MINOS detector coordinate system.  z is along the beam axis.}
\end{center}
\end{figure}

 Since MINOS studies a beam with many more neutrinos than antineutrinos,
the magnet polarity is chosen to focus $\mu^-$ coming from 
the Fermilab direction.
A coordinate system along this beamline
and detector is shown in Figure~\ref{fig:det}.
The z axis is the central axis of the detector, which is also the 
approximate direction of the Fermilab neutrino beam.
\par MINOS has measured the cosmic ray muon charge ratio underground 
\cite{bib:mufson}.  MINOS can only measure the charge of muons that
have less than the maximum detectable momentum (MDM) at the detector.  Above 
that momentum, muon tracks are too straight to measure charge. 
For tracks which do not exit the detector in z,
the MDM only depends on the impact parameter b, which is
the distance of closest approach to the axis (z axis) of the magnet, and
$\theta_z$, the angle with respect to the z-axis.
As discussed in Reference \cite{bib:mufson}, the full magnetic field map, reconstruction
software and detector geometry are used to track muons.  In this note,
certain simplifying assumptions have been made to qualitatively
illustrate certain features of the detector and the MDM.

\section{The calculation of the Maximum Detectable Momentum}
We define the 
Maximum Detectable Momentum (MDM) as that momentum for which a 
nearly straight real track will have a measured curvature (determined
from a fit to points along the track) which is one standard deviation
from zero.  The MDM is simply the reciprocal of the error (s.d.) of
the curvature measurement, when the curvature is expressed in
$(GeV/c)^{-1})$.  The essential features of the magnetic response to cosmic
rays for the MINOS detector can be characterized by considering a
detector which is a right circular cylinder of radius 4m and length
29 m divided into two supermodules.  The radiation length, for 
multiple coulomb scattering, is 3.83 cm.  The magnetic field is
taken to be uniform (in the steel and air), azimuthal and 0.6 tesla.  
All throughgoing tracks which do
not enter or leave
the front or back of the detector curve one way for half of
their trajectory in MINOS and the other way for the other half.  This
will be referred to as an S shaped track.  More than 95\% of cosmic
ray muons in MINOS are S shaped tracks.  This contrasts with charged tracks
along the beam direction, which are C shaped, as long as they don't
cross the center.  The relevant curvature is thus measured twice
along half of the track length.  The approximate length of a half-track is:
\begin{equation}
L = \sqrt{(R^2-b^2)/sin(\theta_z)}
\end{equation}
where R = 4 m is the detector Radius, b is the impact parameter and varies
from -4 m to 4 m, and $\theta_z$ is the angle along the z axis or beam 
axis (it is not the cosmic ray zenith angle).  
\par To calculate the precision
of momentum measurement, we need the component of the magnetic field
which is perpendicular to the track direction.  We approximate this
by finding {\bf B} at the midpoint of the half-track and then resolving
it into components parallel and perpendicular to the track.  It is
\begin{eqnarray}
\begin{tabular}{l}
$B_{perp} = B \times $\\
$\sqrt{[1-\sin^2(\theta_z) \times \cos^2
(arctan(\sqrt{(R^2-b^2)}/2b))]} $\nonumber
\end{tabular}
\end{eqnarray}
Note that both L and $B_{perp}$ depend only on b and $\theta_z$.

\par
\begin{figure}[htbp]
\begin{center}
\includegraphics*[width=6.5cm,angle=0,clip]{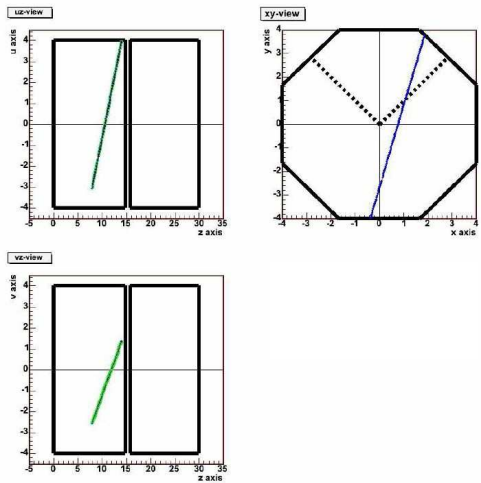}
\caption{\label {fig:view3} 
Three views of a cosmic ray muon data event in MINOS.  Even though this
muon curves more than most cosmic rays in MINOS, the curvature
is not apparent in these views.  Hits which are not in the track
fit are not shown.  The track has a fit momentum of 50.4 $\pm$ 10.2 GeV/c,
and a $\chi^2/$ndof = 119/97.  The charge confidence is about 5$\sigma$.}
\end{center}
\end{figure}

\begin{figure*}[htpb]
\begin{center}
\includegraphics*[width=8.5cm,angle=0,clip]{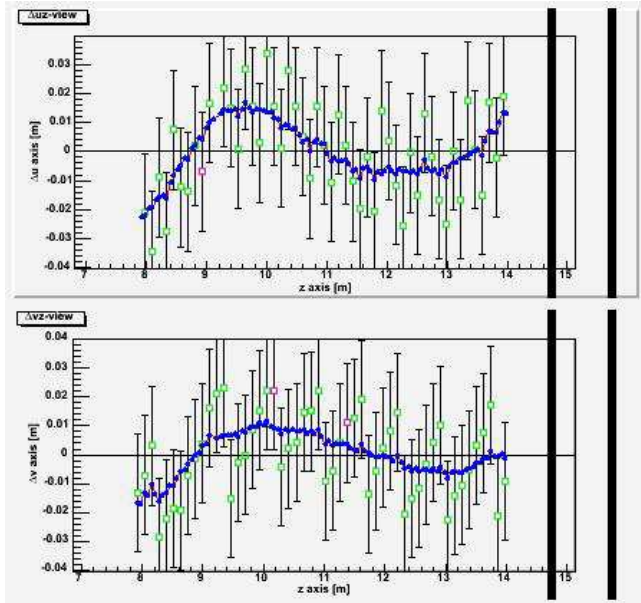}
\caption{\label{fig:view5} 
This shows the deviations from straightness in $\Delta u$
and $\Delta v$
of the cosmic ray muon in Fig. \ref{fig:view3}.  Deviations from
a straight line fit were everywhere less than 2 cm.  
The green squares represent the
centers of the 4 cm wide scintillator strips that had hits
on the track fit.  The vertical lines represent the width
of the strip in u or v.  The solid blue symbols represent the 3D
position of the track fit.  The "S" shape in both views is
indicative of bending in the MINOS toroid and a good track
fit.}
\end{center}
\end{figure*}

 A fairly typical cosmic ray muon in the MINOS detector is shown 
in Figures \ref{fig:view3} and \ref{fig:view5}.
Figure \ref{fig:view3} shows the 3 sides of the detector, uz, vz and xy
with units in meters.  The individual plane crossings are too small
to distinguish in Fig. 2.  To see the curvature we compare the hits to a straight
line fit.  $\Delta u$ and $\Delta v$ are plotted versus z in Figure
\ref{fig:view5}.
Shown are the centers of the
4.1 cm strips, the strips themselves, and the best fit.  Note
that the largest deviations are only about 2 cm, which is comparable
to the strip width.  This muon deviates from straight by almost 5 sigma,
while most cosmic muons have a substantially smaller deviation from straightness.
The largest deviation from the straight line is termed the sagitta of
the half track, and for this track is about 2 cm.

The MDM calculated here is tabulated in Table \ref{fig:mdm1} as a function
of $\theta_z$ and impact parameter, for cosmic ray muon tracks
in a single supermodule.  For the track in Figs 2 \& 3, the MDM is
about 200 GeV/c.

\section{Implications}
The MDM has been useful in reducing the two different kinds of
charge misidentification which provide systematic errors in the measurement
of the atmospheric muon charge ratios.  We call these errors bias errors
and randomization.  Bias errors can have any effect on the measured
charge ratio.  They could be due, for example,
to the acceptance, poor modeling of
the magnetic field or alignment errors.  Acceptance effects can
be accurately studied in a Monte Carlo, but modeling errors cannot.  
Bias
errors do cancel, however, by taking the geometric mean
of the charge ratio from forward and reverse running of the magnetic
field.  Randomization errors are those in which the charge of the
muon is assigned at random.  These always cause the measured $\mu$ charge
ratio to be closer to unity than the true charge ratio, and since
the charge ratio $r>1$, they make the measured r lower.  
They cannot be canceled by using reverse field data.

The curvature resolution is the inverse of the MDM.  The MDM does
not account for bias which must be dealt with separately.
Before forward and reverse field data were combined, there was evidence
for several unexplained bias effects in the MINOS data.  One such
effect is a bump in the charge ratio versus momentum.  We have calculated that
a bias in the sagitta of 2 mm, corresponding to a bias in the curvature
calculation of 1/(2000 GeV/c) together with a curvature resolution of
1/(200 GeV/c) leads to a bump in the charge ratio versus momentum distribution
similar to that seen in our data.  When forward and reverse field were
combined, this bias was completely removed.

We expected randomization effects at large muon momentum above the MDM.
These were minimized by cutting on the charge confidence parameter
$\frac{1}{p}/\sigma(\frac{1}{p})$.  The
charge confidence parameter is approximately equal to the
MDM divided by the measured momentum.
\par We also encountered another randomization
effect at low measured momentum.  This came from an unexpected source
and can be explained as follows.  In Table \ref{fig:mdm2}, the
ratio of the mean
angular deflection
from Multiple Coulomb Scattering (MCS) to the angular deflection from 
bending is calculated
as $0.232/[B_{perp} \times \sqrt{L}]$ with B in tesla and L
in meters.  When this fraction approaches
unity, a straight track will give a good fit to any momentum.  Then
a few hits which should not be fit to the track may be 
incorrectly included in the fit, and
will usually give a low momentum with a random sign of
charge.  This happened
most frequently for directions corresponding to large values in
Table \ref{fig:mdm2}.  These events
were eliminated by requiring track directions inconsistent with this
possibility.

\begin{table}[thpb]
\begin{tabular}{|r|c|c|c|c|c|c|c|c|} \hline
b & 0.0 & 0.5 & 1.0 & 1.5 & 2.0 & 2.5 & 3.0 & 3.5 \\ \hline
$\theta$ & & & & & & & & \\
10 & .14 & .14 & .14 & .14 & .14 & .15 & .15 & .15 \\
20 & .14 & .14 & .15 & .15 & .15 & .15 & .15 & .17 \\
30 & .14 & .14 & .15 & .15 & .16 & .17 & .19 & .22 \\
40 & .15 & .16 & .16 & .18 & .19 & .21 & .24 & .28 \\
50 & .17 & .17 & .18 & .20 & .22 & .25 & .29 & .36 \\
60 & .18 & .18 & .20 & .22 & .26 & .30 & .36 & .47 \\
70 & .19 & .19 & .21 & .24 & .29 & .35 & .45 & .64 \\
80 & .19 & .20 & .22 & .25 & .31 & .39 & .54 & .88 \\ \hline

\end{tabular}
\caption{\label {fig:mdm2} 
Relative angle change from Multiple Coulomb
scattering versus bending
as a function of angle from the z axis (in degrees) and
the absolute value of the impact parameter (in meters).}
\end{table}

\section{Acknowledgments}
This work is supported by US DOE and NSF, 
the UK Particle Physics and 
Astronomy Research Council, and the University of Minnesota.  
We 
wish to thank the Minnesota Department of Natural Resources for use 
of the facilities of the Soudan Underground State Park, and 
also the large crews of workers who helped construct the detector 
and its components, and the mine crew for help in operating the detector.

\begin{table*}[thpb]
\begin{center}
\begin{tabular}{|r|c|c|c|c|c|c|c|c|} \hline
b & 0.0 & 0.5 & 1.0 & 1.5 & 2.0 & 2.5 & 3.0 & 3.5 \\ \hline
$\theta_z$ & & & & & & & & \\
10 & 471 & 470 & 469 & 468 & 467 & 466 & 465 & 464 \\
20 & 471 & 469 & 465 & 460 & 455 & 451 & 447 & 246 \\
30 & 471 & 467 & 458 & 447 & 391 & 298 & 197 & 92 \\
40 & 344 & 333 & 304 & 262 & 213 & 159 & 103 & 47 \\
50 & 235 & 226 & 203 & 171 & 135 & 98 & 62 & 28 \\
60 & 232 & 223 & 197 & 162 & 124 & 87 & 52 & 22 \\
70 & 231 & 221 & 193 & 156 & 115 & 77 & 44 & 17 \\
80 & 270 & 257 & 223 & 177 & 128 & 82 & 43 & 15 \\ \hline
\end{tabular}
\caption{\label {fig:mdm1} 
Maximum detectable momentum (in GeV/c) in MINOS
as a function of angle from the z axis (in degrees) and
absolute value of the impact parameter (in meters).}
\end{center}
\end{table*}

\end{document}